\documentclass[12pt]{article}
\usepackage{amsmath}
\usepackage{graphicx,psfrag,epsf}
\usepackage{enumerate}
\usepackage{natbib}
\usepackage{multirow}
\usepackage{afterpage}
\usepackage{rotating}
\usepackage{url} 
\usepackage{longtable}
\usepackage{caption}

\newcommand{\blind}{1}

\newtheorem{proposition}{Proposition} 
\newtheorem{definition}{Definition} 

\begin{document}

\def\spacingset#1{\renewcommand{\baselinestretch}%
{#1}\small\normalsize} \spacingset{1}


\if1\blind
{
  \title{\bf The Importance of Being Clustered: Uncluttering the Trends of Statistics from 1970 to 2015}
  \author{Laura Anderlucci, Angela Montanari and Cinzia Viroli\\
    Department of Statistical Sciences, University of Bologna, Italy}
    \maketitle
} \fi

\if0\blind
{
  \bigskip
  \bigskip
  \bigskip
  \begin{center}
    {\LARGE\bf The Importance of Being Clustered: Uncluttering the Trends of Statistics from 1970 to 2015}
\end{center}
  \medskip
} \fi

\bigskip
\begin{abstract}
In this paper we retrace the recent history of statistics by analyzing all the papers published in five prestigious statistical journals since 1970, namely: \emph{Annals of Statistics}, \emph{Biometrika}, \emph{Journal of the American Statistical Association}, \emph{Journal of the Royal Statistical Society, series B} and \emph{Statistical Science}.
The aim is to construct a kind of ``taxonomy'' of the statistical papers by organizing and by clustering them in main themes. In this sense being identified in a cluster means being important enough to be uncluttered in the vast and interconnected world of the statistical research.
Since the main statistical research topics naturally born, evolve or die during time, we will also develop a dynamic clustering strategy, where a group in a time period is allowed to migrate or to merge into different groups in the following one. Results show that statistics is a very dynamic and evolving science, stimulated by the rise of new research questions and types of data.
\end{abstract}

\noindent%
{\it Keywords:}  model-based clustering, cosine distance, textual data analysis.
\vfill

\newpage
\spacingset{1.45} 
\section{Introduction}
\label{sec:intro}
It is hard to date the birth of statistics as a modern science. Certainly, in the past forty-five years, remarkable new ideas and contributions to a rich variety of topics were stimulated by the rise of new research questions and new types of data, and disseminated by a wider access to highly-performing electronic devices and scientific journals.

In this work we retrace the recent history of the adult-stage of statistics by analyzing the contributions published in some of the most prestigious statistical journals from 1970 to 2015.

Since classification into distinct entities is a fundamental step to discover meaningful information and to create new knowledge, we aim at constructing a ``taxonomy'' of the considered statistics papers by organizing and clustering them according to main topics. Since the topics of research are many, heterogeneous and they evolve over time, we also develop a dynamic clustering strategy: a group in a decade can migrate or merge into different groups in the following decade; the birth and the (potential) death of topics are allowed as well.
 Of course, it is very hard to disentangle all the possible sub-fields of the statistical research. Suppose a certain number of topics are identified in a period of time: despite their unavoidable degree of internal heterogeneity and their mutual linkage, paraphrasing the title of our work, being clustered is important because it means being uncluttered in the vast and interconnected world of statistics. In other words, a cluster identifies an aggregation of related papers around a relevant statistical topic. In so doing, we assume that a statistical paper is developed around a single research topic. Although we believe that in few cases it can be a restrictive assumption, this unique association is a fundamental condition to create a clear taxonomy of the most important research themes.

Information about the papers is collected as textual data, from their titles and abstracts, and it is stored in a high-dimensional document-term matrix.
Statistical models for analyzing textual data have been mainly developed in the context of information retrieval, natural language processing and machine learning. Proposals aimed at extracting topics from a corpus of documents that consider a unique association between topics and documents use mixtures of specific distributions. The mixtures of unigrams models \citep{Nigam} represent the natural parameterization, because it is based on the idea of modeling the word frequencies as multinomial distributions. However, several authors have shown the superiority of mixtures of von Mises-Fisher distributions for text classification \citep[see][]{Zhong,Banerjee} provided that the textual data are \emph{directional}, that means the frequencies of the documents are normalized to 1 according to the L2 norm.
More sophisticated versions that consider multiple-topic documents exist, namely the latent semantic indexing \citep{Deerwester}, the probabilistic latent semantic models \citep{Hofmann}, Latent Dirichlet Allocation Model \citep{Blei,Blei2,Sun} and more recent elaborated proposals based on graphical models to incorporate information about the co-authorship network \citep[see, for instance,][and references therein]{Bouveyron}.

Embracing the setting of a single association topic-document, we propose an alternative mixture model that overcomes some challenges and computational problems arising from the data characteristics. The proposal stems from the definition of a distance-based density and it is equivalent to the von Mises-Fisher mixture model when data are directional and, likewise the latter, it suffers from the problem of estimating the precision parameter. The model is presented in Section 3, together with a strategy to get a reasonable level of the precision; model estimation is also described. In Section 4 results on the classification of the considered statistical papers are presented for the whole period of time 1970-2015. Finally, in Section 5, we extend the proposed model in a dynamic fashion through a semi-supervised mixture model, in order to describe the evolution over time of the recent scientific research in statistics.

\section{The Data: Statistics in 1970-2015}
\subsection{Data Collection}
The study is based on the articles published on five top statistical journals during 1970-2015: the Annals of Statistics (AoS), Biometrika (Bka), Journal of the American Statistical Association (JASA), Journal of the Royal Statistical Society, series B (JRSSB) and Statistical Science (SSci). These journals have been selected both for their historical importance and for their highest citation metrics in terms of Article Influence Score (AIS) and 5-year Impact Factor among all the statistical journals ranked by the ISI Web of Knowledge database.

More precisely, we considered information contained in titles and abstracts of all the articles published in these journals starting from 1970 or from the first available issue (dated 1973 for AoS and 1986 for SSci). Data have been collected by downloading the freely available bibliography files from the journal websites in RIS format for Statistical Science and in BIB format for the other four journals. Then, the bibliography files have been imported in \verb"R" by using the package \verb"RefManageR" in order to produce a single textual file for each article containing only the title followed by the abstract. Author names and the other editorial information were not considered.
Since our aim is to identify the most relevant topics in the statistical research, we excluded from the analysis the editorial frontmatter articles, the book reviews, the series of papers entitled  ``\emph{A conversation with...}'' published in Statistical Science during the whole period (1986-2015), the interviews with authors narrating career and life rather than statistics and the series ``\emph{Studies in the History of Statistics and Probability}'' published in Biometrika in the first decades. We also excluded the discussion or comments to the articles, replies and rejoinders when they were not accompanied by an abstract. Overall, we collected information on 15472 articles, which are summarized in Table \ref{tab:tabone} by journal and five periods of time: 1970-1979, 1980-1989, 1990-1999, 2000-2009 and 2010-2015.

\begin{table}
\small
\caption{Number of statistical papers published in the five journals by period of time.\label{tab:tabone}}
\begin{center}
\begin{tabular}{lrrrrrr}
Journals & 1970-1979 & 1980-1989 & 1990-1999 & 2000-2009 & 2010-2015 & \textbf{1970-2015}\\\hline
AoS & 610 &	864	&920&	809&	508&	\textbf{3711} \\
Bka & 738&	722&	613&	753&	425&	\textbf{3251} \\
JASA & 1431&	1218&	1393&	1123&	744&	\textbf{5909} \\
JRSSB & 356&	395&	498&	467&	234&	\textbf{1950}\\
SSci       & 0&	69&	151&	239&	192&	\textbf{651} \\
\hline
\textbf{Total} & \textbf{3135}&	\textbf{3268}&	\textbf{3575}&	\textbf{3391}&	\textbf{2103}&	\textbf{15472}\\
\end{tabular}
\end{center}
\end{table}

\subsection{Data Management}
The 15472 textual files were imported in \verb"R" with the library \verb"lsa" so as to produce a document-term matrix containing the term frequencies of each paper.
Raw data were processed by stemming in order to reduce inflected or derived words to their unique word stem. Moreover, we removed numbers and we filtered the terms by a list of English stopwords, that includes the most common words in English, such as adverbs and articles. The whole procedure was automatically performed by using the options available in the \verb"R" function \verb"textmatrix" (library \verb"lsa").
In addition to the default stopwords of the package, we added a list of generic words that are not generally common in English, but that are certainly widespread in the statistical language, such as `variable', `statistics', `analysis', `data' and `model'.
At the end of this filtering process we ended up with a final document-text matrix of 15472 rows, corresponding to the papers, and of 15036 columns, corresponding to the final reduced stemmed terms.

In order to measure the importance of a term in the whole collection of documents, we have weighted each frequency by the so-called Inverse Document Frequency (IDF), which is the logarithm of the total number of documents divided by the number of documents where each term appears. This commonly used normalization \citep{Salton} allows to give more importance to the terms that are contained in the documents but are in general rare.

\section{Clustering Statistics} \label{sec: method}

The basic idea of this work is to cluster papers according to their weighted term frequencies, in order to identify the main relevant topics of Statistics since 1970. Obviously, it is very hard to disentangle all the possible sub-fields of the statistical research. Statistical topics are many, they are naturally interconnected and they evolve over time. However, when, a certain number of clusters, say $k$, are identified, they certainly aggregate similar subtopics of the research. In other words, a cluster identifies an aggregation of related papers around a broad statistical theme and we assume clusters identify the main relevant topics. The internal degree of heterogeneity will depend on the total number of groups $k$.

\subsection{Mixtures of cosine distances}
Mixture models allow to decompose an heterogeneous population into a finite number of sub-groups with an homogeneous density function \citep{Fraley,Peel}. In our case, modeling the component densities of term frequencies is a hard task due to the peculiar characteristics of the data. Each document is characterized by a high-dimensional vector of not-independent term frequencies with a relevant degree of sparsity. The natural model for identifying the topics is the mixture of unigrams models \citep{Nigam}, which is essentially a mixture of multinomial distributions estimable by an EM-algorithm. However, although in general it is an efficient estimation model, results on our (big) data are seriously affected by the amount of zeros and they are extremely sensitive to the initialization of the EM algorithm, which, very frequently, leads the algorithm to be trapped in local unsatisfactory points after very few iterations.

As an alternative, a zero-inflated distribution could be employed to model sparsity. We investigated mixtures of zero-inflated Poisson, Bernoulli and Negative Binomials, but such choices did not offer a satisfactory approximation to the observed distributions for two principal reasons. Firstly, the theoretical zero-inflated models involve a very large number of parameters, since (at least) two different parameters have to be estimated for each term and each group, namely the zero-inflation and the location parameters. As a consequence, the fit is computationally unfeasible. Secondly, these univariate distributions would be fitted to each observed set of term frequencies, interpreted as a variable, therefore ignoring the semantic dependence among the terms.

Due to these difficulties we changed our perspective from density based estimation to distance-based clustering models. Distance-based densities have been successfully used by several authors \citep[see][]{Ma57,FV86,Di88} in the context of ranking data and then adapted for classification in a mixture-based perspective by \cite{MM03}. These models can be viewed as special cases of the so-called `probabilistic D-clustering' \citep{israel}. In the context of ranking data, several distance measures have been used, e.g. Euclidean, Kendall, Spearman and Cayley’s distances. None of them provides a useful measure for sparse textual data, since they may be highly affected by the high proportion of zeros. A prominent measure of distance overcoming these difficulties is based on the cosine similarity, because it considers only the non-zero elements of the vectors, allowing to measure the dissimilarity between two documents in terms of their subject matter. Given two $p$-dimensional documents, say $\textbf{x}$ and $\textbf{y}$, the cosine distance of the two corresponding frequency vectors is:

\begin{eqnarray}\label{eqn:cosdist}
d(\textbf{x},\textbf{y})=1-\frac{\sum_{h=1}^px_hy_h}{\sqrt{\sum_{h=1}^px_h^2}\sqrt{\sum_{h=1}^py_h^2}},
\end{eqnarray}

where $x_h$ and $y_h$ denote the frequency of word $h$ in document $\textbf{x}$ and $\textbf{y}$, respectively. This measure is not affected by the amount of zeros and is a normalized synthesis of the $p$-variate terms of the documents. Since the elements of $\textbf{x}$ and $\textbf{y}$ are positive or null frequencies, it is easy to prove that the distance ranges between 0 and 1.

Given the cosine distance $d(\textbf{y},\boldsymbol\xi)$ of a generic document $\textbf{y}$ from a reference centroid, say $\boldsymbol\xi$, we define a probability density function for the random variable $\textbf{y}$ as
\begin{eqnarray}\label{eqn:vonMises}
 f(\textbf{y};\boldsymbol\xi,\lambda)=\psi(\lambda)e^{-\lambda d(\textbf{y},\boldsymbol\xi)}
\end{eqnarray}

where $\lambda$ is a positive precision (or concentration) parameter, with $\lambda > 0$, and $\psi(\lambda)$ is a normalization constant such that $f(\textbf{y};\boldsymbol\xi,\lambda)$ is a proper density function. When $\textbf{y}$ and $\boldsymbol\xi$ are distributed on the surface of a unit hypersphere, so that they are directional, the density in (\ref{eqn:vonMises}) is the von Mises-Fisher distribution and its normalization constant analytically exists as a function of the modified Bessel function of the first kind and order $p/2-1$ \citep{Mardia}. Mixtures of von Mises-Fisher distributions have been largely used by many authors in the information retrieval community for clustering direction data under the assumption that the direction of a text vector is more important than its
magnitude \citep[see, for more details,][]{Banerjee,Peel,Zhong}.

In our data problem, many words are removed as either stopwords or very widespread statistical terms. Moreover, as it will be explained in Section 4, the analysis will be performed on a subset of selected variables, chosen according to their entropy in order to remove biases due to the high-dimensionality and to the presence of irrelevant words. In this perspective, data cannot be normalized into directional data and, therefore, analyzing the absolute values of the frequencies is preferable. In order to perform clustering, we consider a mixture of $k$ cosine distance density functions:
\begin{eqnarray}\label{eqn:mixture}
f(\textbf{y};\boldsymbol\xi,\lambda)=\sum_{i=1}^k \pi_i \psi(\lambda)e^{-\lambda d(\textbf{y},\boldsymbol\xi_i)}
\end{eqnarray}
with positive mixture weights $\pi_i$, summing to unity, $\sum_{i=1}^k \pi_i=1$, and component varying centroid vectors $\boldsymbol\xi_i$. Notice that in this case the normalization quantity $ \psi(\lambda)$ cannot be derived in closed form, thus making the estimation of $\lambda$ hard. In the next Section we will show a strategy to get a reasonable value for the precision parameter $\lambda$.

\subsection{Role of the precision parameter}
The precision parameter $\lambda$ is taken common among the mixture components for theoretical and practical reasons. Firstly, observe that the precision acts as a scaling of the distances. For high values of $\lambda$ even small differences between distances induce relevant differences in the density values. In this case, a small difference between the cosine distance of a document $\textbf{y}$ from two centroids, say $\boldsymbol\xi_1$ and $\boldsymbol\xi_2$, is over-weighted by $\lambda$ and it produces a relevant difference in terms of likelihood, favouring the posterior clustering to the component with smallest distance. On the contrary, when $\lambda \rightarrow 0$ the importance of the distances vanishes and the densities converge to the uniform distribution. In other terms, as $\lambda$ increases, the mixture is forced to produce more homogeneous and ``purer'' clusters (in order to have a good fit this implies to have more components). Fixing $\lambda$ across components implies all clusters have the same degree of internal homogeneity.

A more theoretical insight about the crucial role of $\lambda$ for clustering derives by imposing a consistency relation between distances and mixture posterior classification. More precisely, for a generic document $y$, let $i'$ denote the component with the minimum distance, i.e. $i' : min_{1\leq i \leq k} \{d (y,\xi_i)\}=d (y,\xi_i')=d_{i'}$. Then the following definitions and propositions establish a formal consistency relation between the value of $\lambda$ and the posterior classification, which justifies the choice of a common dispersion parameter for all the mixture components.

\begin{definition}
Given the mixture model (\ref{eqn:mixture}), the \emph{consistent clustering rate}, say $1-\alpha$, is the probability of allocating $y$ to the component to which it has the minimum distance:
$$1-\alpha=Pr(z_{i'}=1|y,d_{i'})$$
where $z$ is the hidden allocation vector of length $k$ with value one in correspondence of the component membership and zero otherwise.
\end{definition}

In a similar manner we may define the inconsistent clustering rate:
\begin{definition}
Given the mixture model (\ref{eqn:mixture}), the \emph{inconsistent clustering rate} is defined as:
$$\alpha=Pr(z_{i'}\neq 1|y,d_{i'})$$
\end{definition}

The following results establish a formal relation between $\lambda$ and the inconsistent clustering rate.
\begin{proposition}
Given $d_i=d(y,\xi_i)$ for $i=1,\ldots,k$ and $i' : min_{1\leq i \leq k} \{d (y,\xi_i)\}=d_{i'}$, the inconsistent clustering rate for the model (\ref{eqn:mixture}) is inversely and non linearly related to $\lambda$ through the formula
\begin{eqnarray}\label{eqn:rel}
\alpha=\frac{\sum_{i=1}^k \pi_i e^{-\lambda (d_i-d_{i'})} - \pi_{i'}}{\sum_{i=1}^k \pi_i e^{-\lambda (d_i-d_{i'})}}
\end{eqnarray}
\end{proposition}
To prove the proposition observe that by definition we have
\begin{eqnarray}\label{eqn:rel2}
1-\alpha=\frac{\pi_{i'}f(y|z_{i'}=1)}{\sum_{i=1}^k\pi_{i}f(y|z_{i}=1)}=\frac{\pi_{i'}\psi(\lambda)e^{-\lambda d_{i'}}}{\sum_{i=1}^k\pi_{i}\psi(\lambda)e^{-\lambda d_{i}}}=\frac{1}{1+\sum_{i\neq i'}\frac{\pi_{i}}{\pi_{i'}}e^{-\lambda(d_i-d_{i'})}}.
\end{eqnarray}
Now equation (\ref{eqn:rel}) derives by observing that $\sum_{i\neq i'}\frac{\pi_{i}}{\pi_{i'}}e^{-\lambda(d_i-d_{i'})}+1=\sum_{i=1} ^k\frac{\pi_{i}}{\pi_{i'}}e^{-\lambda(d_i-d_{i'})}$.
Generally, as $\lambda$ increases $\alpha$ decreases and viceversa. More formally,

\begin{proposition}
Given the relationship (\ref{eqn:rel}):
\begin{eqnarray}\label{eqn:rel3}
 \lim_{\lambda \rightarrow \infty} \alpha(\lambda)=0, \ \ \ \ \ \ \lim_{\lambda \rightarrow 0} \alpha(\lambda)=1-\pi_{i'}.
 \end{eqnarray}
\end{proposition}

The first limit derives by observing that $\sum_{i=1}^k \pi_i e^{-\lambda (d_i-d_{i'})}=\pi_{i'}+ \sum_{i \neq i'}^k \pi_i e^{-\lambda (d_i-d_{i'})} \geq \pi_{i'}$ for all $\lambda>0$ because $(d_i-d_{i'})> 0$ for $i\neq i'$. The second limit derives directly from the right-hand part of equation (\ref{eqn:rel2}).

\begin{figure}[h]
  \centering
\includegraphics[scale=0.7]{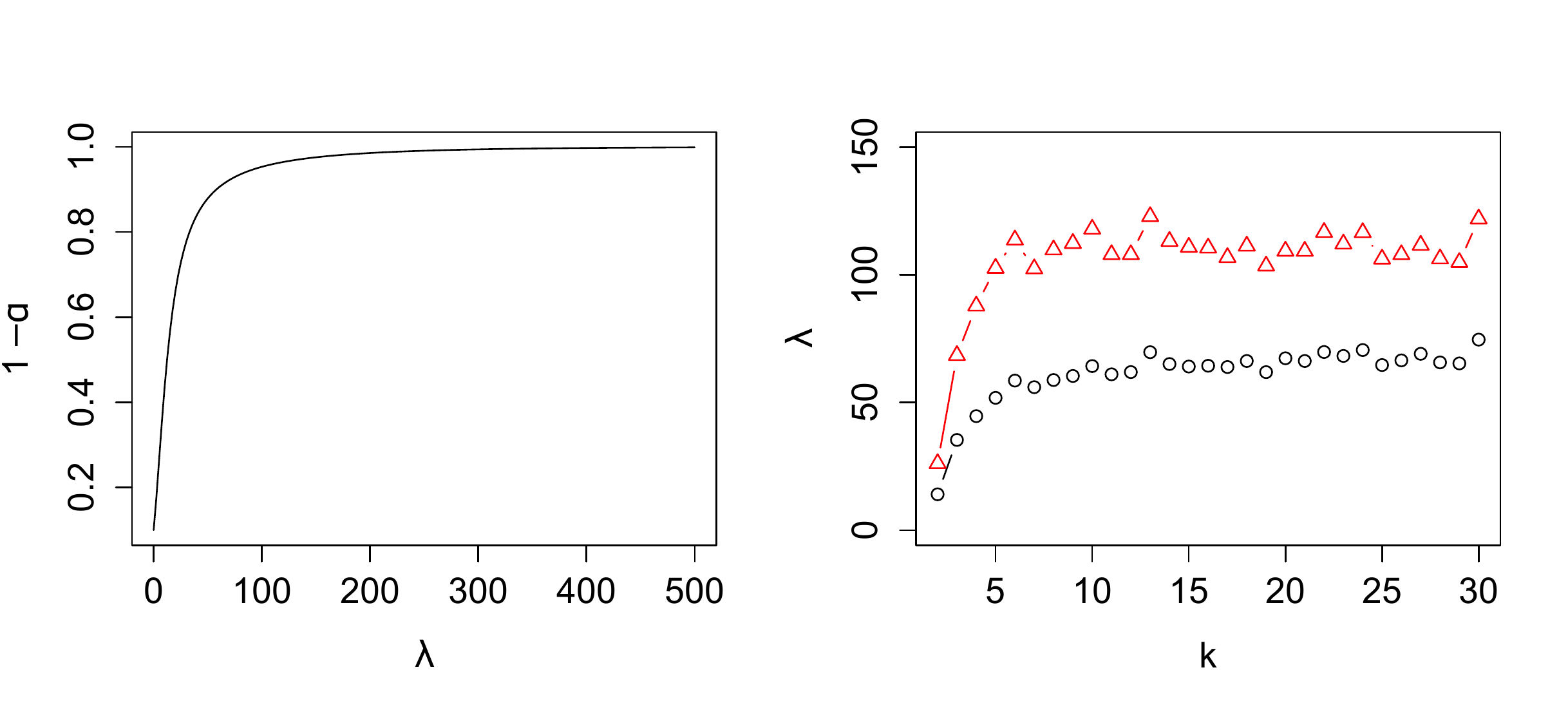}
  \caption{The effect of $\lambda$: in the first panel the relation between $\lambda$ and $1-\alpha$ is shown; the second panel shows the value of $\lambda$ corresponding to  an average consistent clustering rate of 0.90 (circles points) and 0.95 (triangle points) as $k$ varies. }\label{fig: lambda}
\end{figure}

Figure (\ref{fig: lambda}) (first panel) shows the relation between $\lambda$ and the average consistent clustering rate in the dataset with all the 15472 documents and $k=10$ components. The second panel of the Figure shows the best value of $\lambda$ to get an average consistent clustering rate of 0.90 (circles points) and 0.95 (triangle points) respectively on the same data as $k$ ranges between 2 to 30.

From these results it is clear that, in order to have the same goodness of fit, a mixture model with higher $\lambda$ would generally require more components. Moreover, the inverse relation between $\lambda$ and the the inconsistent clustering rate in (\ref{eqn:rel}) can be used to approximate the precision parameter as explained in the next Section.

\subsection{Model estimation} \label{subsec:modelestimation}
A computational problem of the mixture model (\ref{eqn:mixture}) is related to the estimation of the normalization constant $\psi(\lambda)$, that cannot be derived neither analytically nor numerically due to the complexity of the cosine distance and to the high-dimension of the multiple integral to be evaluated. This translates into the problem of estimating the precision parameter $\lambda$. According to Proposition 1, a way to get a reasonable value for the precision parameter is to fix a desirable level for the consistent clustering average rate among all the observations, such as $1-\alpha=0.95$. For a given value of $\alpha$, $\lambda$ can be obtained by solving equation (\ref{eqn:rel}) over the sum of all the observations. The other parameters of the mixture model (\ref{eqn:mixture}) can be easily estimated via a conventional EM algorithm by maximizing the complete-data log-likelihood. We denote by $y_{j}=(y_{j1},\ldots,y_{jp})$ the vector of weighted IDF frequencies of the $j$th ($j=1,\ldots,n$) document and by $z_{j}=(z_{j1},\ldots,z_{jk})$ the latent allocation variable, that takes the value 1 in correspondence of the cluster membership and zero otherwise. The complete log-likelihood is:
\begin{eqnarray}\label{eqn:logL}
\ell_C(\boldsymbol\xi,\boldsymbol\pi;\textbf{y},\textbf{z})
=\sum_{j=1}^n\sum_{i=1}^kz_{ji}\left(\log \pi_i + \log \psi(\lambda)-\lambda d(y_j,\xi_i)\right)
\end{eqnarray}
The two steps of the EM algorithm are:

\emph{E-Step:}\\
Compute the posterior probabilities as
$$\hat{z}_{ji}=\frac{\pi_i e^{-\lambda d(y_j,\xi_i)}}{\sum_{i=1}^k\pi_i e^{-\lambda d(y_j,\xi_i)}}$$

\emph{M-Step:}\\

\noindent (a) Compute $\lambda$ by solving $$\sum_{j=1}^n\sum_{i=1}^k\frac{\pi_{i}}{\pi_{i'}}e^{-\lambda(d_{ij}-d_{i'j})}=\frac{n}{1-\alpha}$$ where $d_{ij}=d(y_j,\xi_i)$.

\noindent (b) Compute via numerical optimization methods the values of the centroids as
$$\xi_{i}=\underset{\xi}{\operatorname{argmin}} \sum_{j=1}^n \hat{z}_{ji} d(y_j,\xi).$$

\noindent (c) Compute the weights as
$$\pi_{i}=\sum_{j=1}^n \hat{z}_{ji} /n.$$

The algorithm converges quickly, but, it is sensitive to its starting values. To avoid to get stuck in local maxima we initialized it with the solution of the spherical $k$-means based on the cosine distance \citep{Inder,Maitra}.

\section{Overview of major clustered Statistical Topics from 1970 to 2015} \label{sec:static_clustering}
The richness of the statistical contributions of the past forty-five years is so broad that no clustering can exhaustively describe the variety of topics and ideas; even if it could, the results would be unintelligible anyway.

The clustering procedure described in Section \ref{sec: method} can help disentangling the principal trends that characterized the statistical research of the last half century.
The idea is to obtain a picture of the top twenty-five topics that have led the research since 1970. As previously stated, it will not be an exhaustive list: it would rather represent the most important twenty-five topics (or, at least, important enough to become separate clusters) that have been discussed in the literature so far.

The five chosen journals are standard reference in the statistical literature, and their articles, given the high number of citations, can be considered representative of the main research trends in Statistics. However, each journal has a very distinctive character and the discussed themes could be different. For this reason we decided to also separately zoom in on the periodicals: distinct analyses allow to identify and distinguish transversal topics, highlighted by every journal, from more specific subjects, that only appeared in a subset.

\subsection{Analysis} \label{subsec:staticresults}
The global document-term matrix is very large (15472 $\times$ 15036) and sparse (i.e. many rare or misspelled words appear in a paper only); these characteristics make it hardly tractable as it is: a variable selection is needed so as to narrow the dimensionality.

\begin{figure}[t]
  \centering
\includegraphics[scale=0.6]{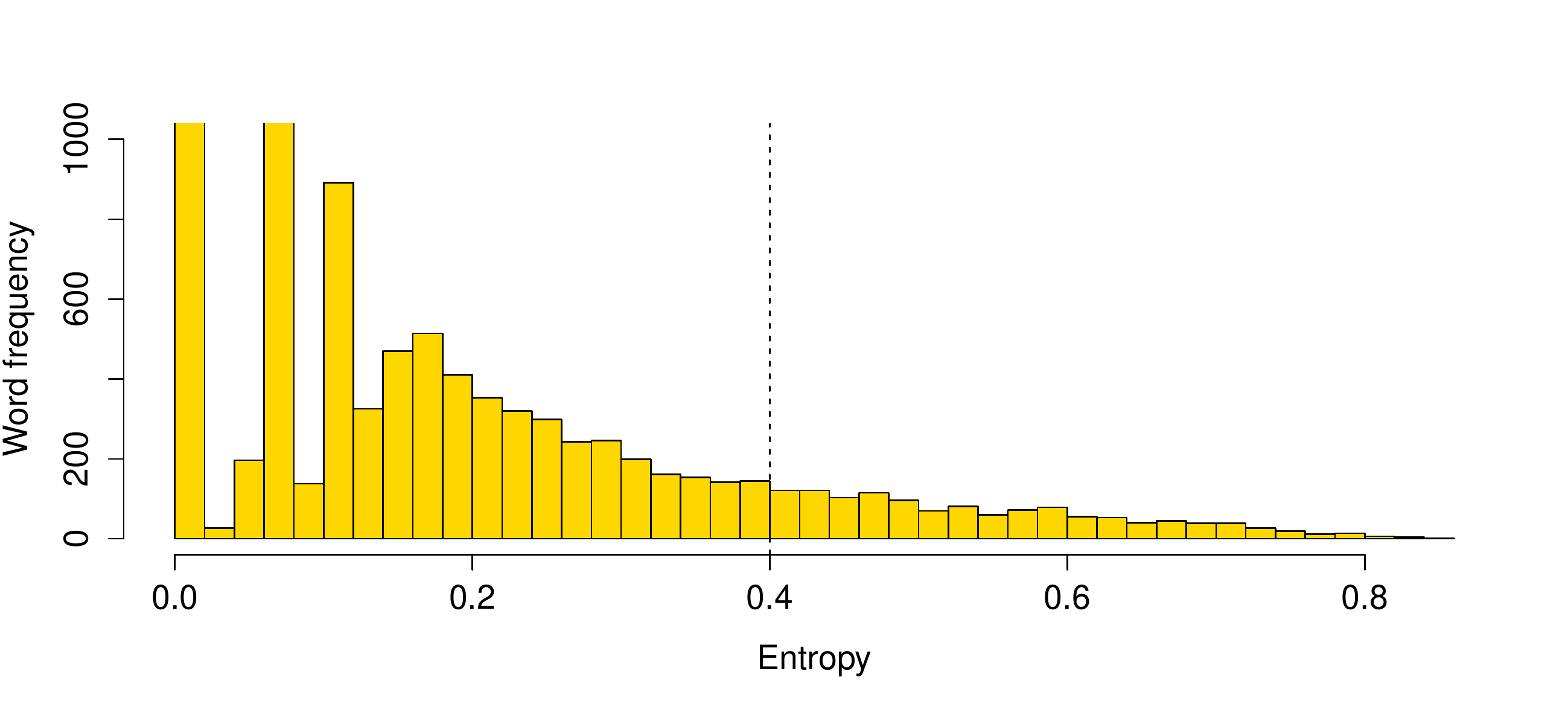}
  \caption{Histogram of the word entropies calculated in the complete dataset. The dashed line denotes a cutoff of 0.4.}\label{fig: entropy}
\end{figure}

We are interested in selecting those words that help identifying homogeneous groups, that is, that are differentially present across the documents. A classic and popular measure, able to capture the \emph{noise} in a single distribution, is \emph{Shannon's entropy} \citep{Shannon}; the entropy $H$ for a term $h$ is calculated as:
\[ H_h=-\sum_{j=1}^n  \frac{f_{jh}\log{f_{jh}}}{\log{n}},  \quad  0\leq H_h \leq 1, \quad h=1,\ldots,p   \]
where $f_{jh}$ is the relative frequency of word $h$ in document $j$, $n$ is the number of documents. Values of $H_h$ close to zero refer to words that are very rare in the considered set of documents; these terms are not informative for clustering and they can also be potentially insidious because of their excess of zeros. Therefore, a natural way to perform variable selection is to set a lower bound for entropy values. By inspecting the entropy distribution of all the words contained in the complete dataset (see Figure \ref{fig: entropy}) we considered only the terms with $H_h\geq 0.40$, it being the approximated middle point between the minimum and maximum observed entropy.

Table \ref{tab:words_journ} contains the number of words for each journal, with and without imposing a constraint on the minimum entropy. Variable selection so performed allowed to deal with much smaller but still meaningful data sets.

\begin{table}
\small
\caption{Number of words for each journal with (first row) and without (second) a threshold on the minimum entropy value $H$.} \label{tab:words_journ}
\begin{center}
\begin{tabular}{lrrrrrr}
\hline \textbf{No. of words}      & \textbf{AoS}   & \textbf{Bka}    & \textbf{JASA}    & \textbf{JRSSB} &\textbf{ SSci} & \textbf{All journals}\\
\hline
$H\geq$0.40             &703    &	632         &1021     &	604&	483 & \textbf{1272} \\
Unconstrained $H$               &5874   &   5460        &10769    &	4561&	4349 & \textbf{15036}  \\
\hline
\end{tabular}
\end{center}
\end{table}

For each of the six datasets, we used the cosine-distance $k$-means algorithm (with $k$=25, and five different runs) as initial values for the EM algorithm. The desirable average consistency rate in (a) step of the estimation algorithm in Section \ref{subsec:modelestimation} was fixed at $1-\alpha=95\%$.

Since the idea was to identify the top twenty-five topics, the number of groups was set, rather than chosen by some information criteria. For this reason, a few abstracts may have ended up being assigned to a group that is meaningfully not close, but that still represents the nearest one. To describe the homogeneity of the clusters we defined and computed a cohesion index $C_i$ as:
\[ C_i=\sqrt{1-\bar{d_i}^2}  \qquad  \qquad 0 \leq C_i \leq 1,\]
where $\bar{d_i}$ is the average cosine distance between all papers within cluster $i$; the closer $C_i$ is to 1, the more homogeneous cluster $i$ is.

\begin{figure}[!h]
\begin{center}
  \includegraphics[scale=0.75]{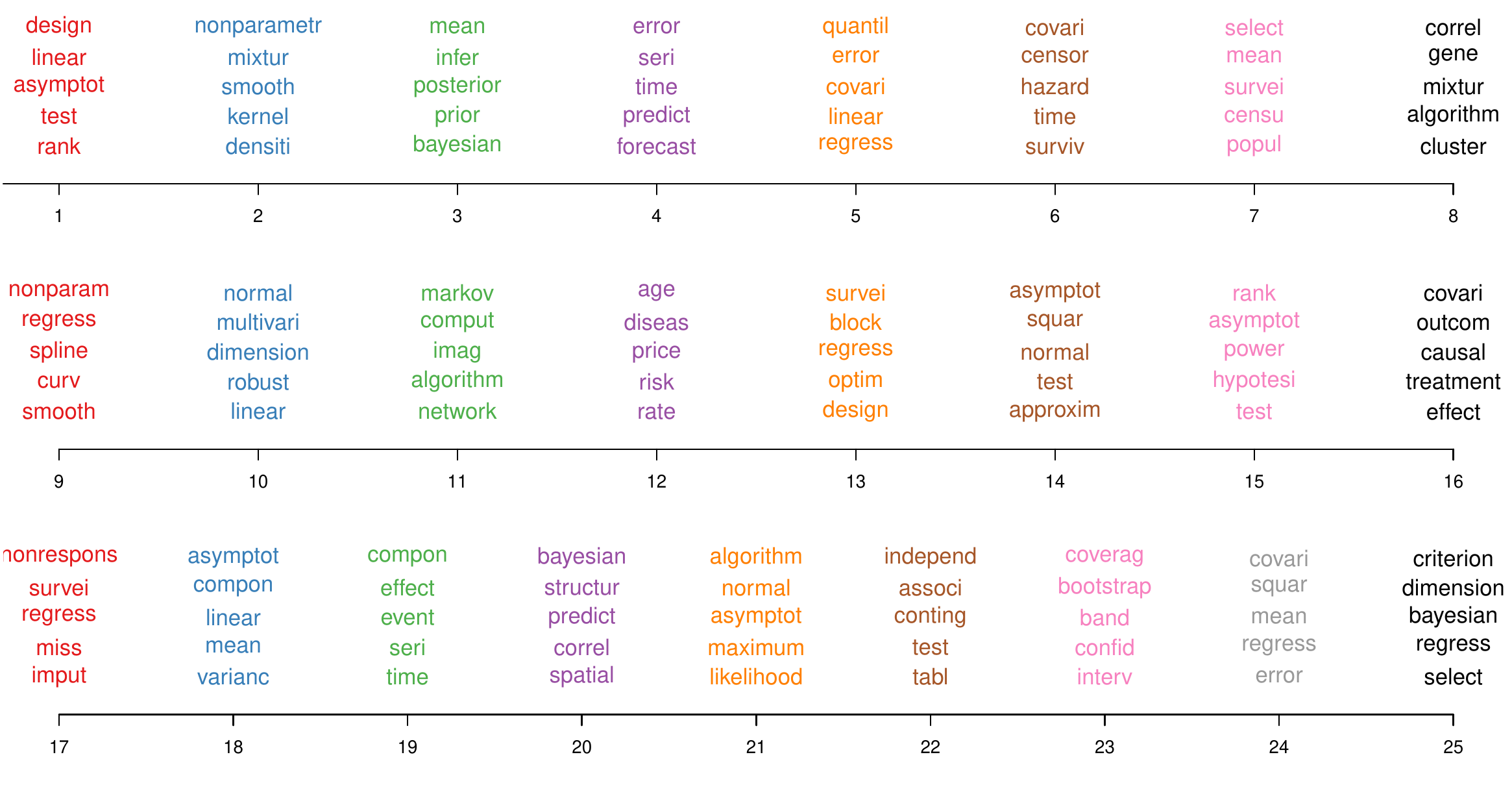}
  \caption{Most frequent words (from bottom to top) for the estimated 25 groups of JASA.}\label{fig: topic.pdf}
  \end{center}
\end{figure}

\subsection{Results}

To fully characterize the estimated groups and to identify the corresponding topics, we considered the five most frequent words (according to the IDF-corrected frequencies) contained in a group and the most representative paper having the minimum cosine distance from the corresponding centroid. Figure \ref{fig: topic.pdf} and Table \ref{tab:rpaper} show the results for the analysis conducted on the papers published in the Journal of the American Statistical Association from 1970 to 2015, both in the sections Theory and Methods and Applications and Case Studies.

\begin{footnotesize}
\begin{longtable}{rp{0.85\textwidth}}
\caption{JASA data: the most representative paper in each group.}
\label{tab:rpaper} \\
\hline
\textbf{Cluster} &\textbf{Paper} \\
\hline
 1 & M.A. Fligner and S\.N. MacEachern, \emph{Nonparametric Two-Sample Methods for Ranked-Set Sample Data}, 2006\\
 2 & A.J. Izenman, \emph{Review Papers: Recent Developments in Nonparametric Density Estimation}, 1991\\
 3 & F.J. Samaniego and A. Neath, \emph{How to be a Better Bayesian}, 1996\\
 4 & A.H. Murphy and R.L. Winkler, \emph{Probability Forecasting in Meteorology}, 1984 \\
 5 & R.E. Barlow and H.D. Brunk, \emph{The Isotonic Regression Problem and its Dual}, 1972 \\
 6 & J.D. Bebchuk and R.A. Betensky, \emph{Local Likelihood Analysis of Survival Data With Censored Intermediate Events}, 2001 \\
 7 & B.C. Arnold and R.A. Groeneveld, \emph{Maximal Deviation between Sample and Population Means in Finite Populations}, 1981 \\
 8 & R. Peck, L. Fisher and J. Van Ness, \emph{Approximate Confidence Intervals for the Number of Clusters}, 1989 \\
 9 & D. Nychka, \emph{Choosing a Range for the Amount of Smoothing in Nonparametric Regression}, 1991 \\
 10 & J.L. Gastwirth and M.L. Cohen, \emph{Small Sample Behavior of Some Robust Linear Estimators of Location}, 1970 \\
 11 & J. Zhang and Y. Chen, \emph{Sampling for Conditional Inference on Network Data}, 2013 \\
 12 & R.A. Holmes, \emph{Discriminatory Bias in Rates Charged by the Canadian Automobile Insurance Industry}, 1970 \\
 13 & B. Jones and D. Majumdar, \emph{Optimal Supersaturated Designs}, 2014 \\
 14 & G.S. Easton and E. Ronchetti, \emph{General Saddlepoint Approximations with Applications to L Statistics}, 1986 \\
 15 & D.I. Tang, \emph{Uniformly More Powerful Tests in a One-Sided Multivariate Problem}, 1994 \\
 16 & M.E. Sobel, \emph{What Do Randomized Studies of Housing Mobility Demonstrate?}, 2006 \\
 17 & R.J.A. Little, \emph{Regression with Missing X's: A Review}, 1992 \\
 18 & P.H. Westfall, \emph{Computable MINQUE-Type Estimates of Variance Components}, 1987 \\
 19 & J.H. Stock, \emph{Estimating Continuous-Time Processes Subject to Time Deformation}, 1988 \\
 20  & D.J. Nordman and P.C. Caragea, \emph{Point and Interval Estimation of Variogram Models Using Spatial Empirical Likelihood}, 2008 \\
 21 & H. Frydman, \emph{Maximum Likelihood Estimation in the Mover-Stayer Model}, 1984 \\
 22 &  L.A. Goodman, \emph{Partitioning of Chi-Square, Analysis of Marginal Contingency Tables, and Estimation of Expected Frequencies in Multidimensional Contingency Tables}, 1971 \\
 23 & P.H. Peskun,\emph{A New Confidence Interval Method Based on the Normal Approximation for the Difference of Two Binomial Probabilities}, 1993 \\
 24 & A.S. Whittemore and J.B. Keller, \emph{Approximations for Regression with Covariate Measurement Error}, 1988 \\
 25 & R.E. Weiss, \emph{The Influence of Variable Selection: A Bayesian Diagnostic Perspective}, 1995 \\
 \hline
\end{longtable}
 \end{footnotesize}

The balloon plot of Figure \ref{fig: balloonJASA.pdf} gives a picture on the top 25 leading topics published in JASA. The size of the balloons is proportional to the group dimension and the colors are shaded by cohesion index, i.e. the lighter the colour, the less homogeneous a cluster is. The cluster position has been computed by multidimensional scaling.

\afterpage{%
\thispagestyle{empty}
\begin{figure}[!h]
\begin{center}
  \includegraphics[scale=0.75,angle=90,origin=c]{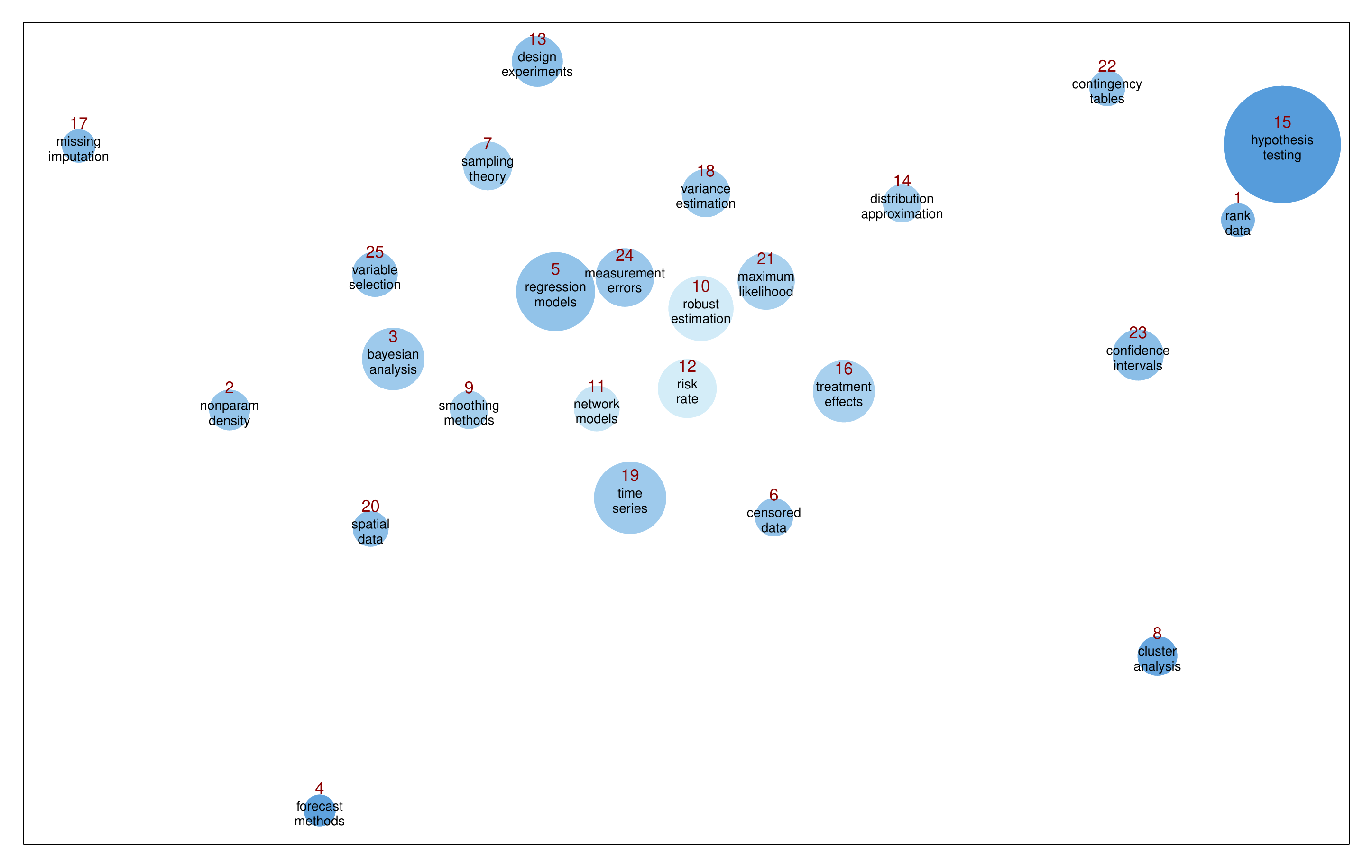}
  \caption{Balloon plot of the top 25 leading topics published in JASA.}\label{fig: balloonJASA.pdf}
  \end{center}
\end{figure}
\clearpage
}

It is not surprising that the topic \emph{hypothesis testing} is the biggest group, while other topics, such as \emph{missing imputation} and \emph{forecast methods} contain less contributions. Some popular research themes are missing. Think at bootstrap, classification or dimension reduction methods. They have been clearly absorbed by the other estimated groups, that in this sense represent the more important broad themes for JASA in the last 45 years. For instance, the major contributions on bootstrap are contained in the \emph{hypothesis testing} and \emph{confidence intervals} groups.
A more extensive view of the research topics of major interest in the statistical literature can be obtained by a complete analysis on all the considered journals.

Table \ref{tab:tot} contains the top 25 leading topics published in the five journals since 1970, sorted by homogeneity. The first column shows the cluster id; the number of abstracts and the cohesion index for each group are reported in third and fourth column, respectively.
Tables \ref{tab:annals}, \ref{tab:biometrika}, \ref{tab:jasa}, \ref{tab:jrssb} and \ref{tab:ss} include similar information for AoS, Bka, JRSSB, JASA and SSci, respectively.

Classic themes (like \emph{hypothesis testing}, \emph{confidence intervals}, \emph{regression models}, \emph{Bayesian analysis}, \emph{design of experiments}, \emph{time series}) formed separate clusters in all the journal-specific analysis. \emph{Hypothesis testing} constitutes the biggest (with more than 2500 papers) and most homogeneous group of abstracts, its relevance is remarkable. Regression and graphical models are fairly big groups, carrying about a thousand papers each.

Other topics in Table \ref{tab:tot} can be found in many of the considered journals; these are broad concepts and of general interest, like - among others -  \emph{estimation algorithms}, \emph{prediction analysis}, \emph{density estimation}, \emph{treatment effect}, \emph{graphical models}, \emph{variable selection}, \emph{bootstrap methods} and \emph{classification}.

\emph{Contingency table}, \emph{dimension reduction}, \emph{asymptotic properties of estimators} are examples of more specific arguments: their contribution is globally remarkable (hundreds of papers with a good degree of cohesion) despite their resonance echoed in only one of the five journals.

Although they did not result in the global clustering, other statistical topics (e.g. \emph{variance estimation}, \emph{robustness}, \emph{spatial statistics}, \emph{missing data}, \emph{networks}) were relevant for JASA and a few other periodicals. The Annals of Statistics has some important clusters on probabilistic and inferential issues, Biometrika on population selection and inference, the JRSSB on generalized linear models and probability; Statistical Science is the only journal where genomic and genome-wide association studies groups appeared, together with clinical trials and, due to its peculiar character, statistics historical papers and Fisher's work reviews.

%
%
%
%

\begin{table}
\caption{Top 25 leading topics published in the five statistical journals (\emph{Annals of Statistics}, \emph{Biometrika}, \emph{Journal of the American Statistical Association}, \emph{Journal of the Royal Statistical Society: Series B} and \emph{Statistical Science}), from 1970 to 2015.} \label{tab:tot}

\centering \footnotesize
\begin{tabular}{rlcc}

  \hline
 \multirow{2}*{\textbf{Cluster}} &\multirow{2}*{\textbf{Topic}} & \textbf{No. of} & \textbf{Cohesion}\\
                       &                    &  \textbf{ papers}   &\textbf{ index }\\
    \hline
  4 & Hypothesis testing & 2516 & 0.82 \\
  9 & Bootstrap methods & 214 & 0.82 \\
 13 & Design of experiments & 912 & 0.77 \\
 19 & Cluster analysis and model-based clustering & 247 & 0.77 \\
  8 & Rank data analysis & 225 & 0.72 \\
 16 & Confidence intervals & 439 & 0.67 \\
 15 & Regression models & 1105 & 0.65 \\
 21 & Density estimation & 477 & 0.65 \\
 25 & Contingency tables & 173  & 0.65 \\
  5 & Maximum likelihood & 970 & 0.64 \\
  1 & Variable and model selection & 397 & 0.63 \\
 20 & Distribution approximation methods & 313 & 0.62 \\
  3 & Bayesian analysis & 917 & 0.61 \\
 12 & Prediction analysis & 354 & 0.60 \\
 17 & Time series & 858 & 0.59 \\
 18 & Measurement error models & 553 & 0.57 \\
 22 & Censored data and survival analysis & 425 & 0.57 \\
 11 & Estimation algorithm: EM, MCMC, ... & 378 & 0.56 \\
 2 & Linear combination and estimators & 487 & 0.55 \\
  6 & Treatment and causal effects & 594 & 0.55 \\
 24 & Dimension reduction & 346 & 0.55 \\
 23 & Asymptotic properties of estimators & 568  & 0.53 \\
  7 & Sampling theory & 455 & 0.52 \\
 10 & Rate of convergence determination & 515 & 0.47 \\
 14 & Graphical models & 1034 & 0.31 \\
  \hline
\end{tabular}
\end{table}

\begin{table}
\caption{Top 25 leading topics published in the \emph{Annals of Statistics}, from 1970 to 2015.} \label{tab:annals}

\centering \footnotesize
\begin{tabular}{rlcc}
  \hline
 \multirow{2}*{\textbf{Cluster}} &\multirow{2}*{\textbf{Topic}} & \textbf{No. of} & \textbf{Cohesion}\\
                       &                    &  \textbf{ papers}   &\textbf{ index }\\
    \hline
  1 & Design of experiments: optimality & 251 & 0.85 \\
 15 & Quantile estimators & 60 & 0.82 \\
 21 & Graphical models & 64 & 0.82 \\
 17 & Bootstrap methodology & 78 & 0.81 \\
 24 & Hypothesis testing & 525 & 0.81 \\
  5 & Models and methods for rank data & 72 & 0.75 \\
  9 & Prediction functions and predictive models & 59 & 0.73 \\
 19 & Priors in Bayesian analysis & 100 & 0.73 \\
  6 & Density estimation & 191 & 0.72 \\
 13 & Bayesian analysis & 127 & 0.72 \\
 20 & Regression models & 276 & 0.70 \\
 25 & Confidence intervals & 105  & 0.68 \\
 12 & Maximum likelihood & 240 & 0.67 \\
  3 & Estimation algorithms: convergence and properties & 72 & 0.63 \\
 16 & Time series & 132 & 0.62 \\
 18 & Rate of convergence determination & 161 & 0.61 \\
 22 & Asymptotic properties of estimators & 279 & 0.61 \\
 23 & Probabilistic lower and upper bounds & 102  & 0.60 \\
  2 & Covariance matrix estimation & 105 & 0.59 \\
 10 & Optimal decision rule and optimality criteria & 134 & 0.59 \\
  7 & Linear models and combinations & 131 & 0.57 \\
  4 & Distribution approximations (saddlepoint, Laplace, ...) & 59 & 0.54 \\
  8 & Exponential family properties & 78 & 0.53 \\
 11 & Admissible minimax and other mean estimators & 148 & 0.50 \\
 14 & Random variable probabilistic results & 162 & 0.44 \\
 \hline
\end{tabular}
\end{table}

\begin{table}
\caption{Top 25 leading topics published in \emph{Biometrika}, from 1970 to 2015.} \label{tab:biometrika}

\centering \footnotesize
\begin{tabular}{rlcc}
  \hline
 \multirow{2}*{\textbf{Cluster}} &\multirow{2}*{\textbf{Topic}} & \textbf{No. of} & \textbf{Cohesion}\\
                       &                    &  \textbf{ papers}   &\textbf{ index }\\
    \hline
 18 & Hypothesis testing & 614 & 0.84 \\
 23 & Design of experiments & 261  & 0.84 \\
  4 & Cluster analysis & 41 & 0.82 \\
  2 & Priors in Bayesian analysis & 77 & 0.77 \\
 15 & ARMA models & 75 & 0.74 \\
 24 & Maximum likelihood & 293 & 0.73 \\
 12 & Prediction analysis & 55 & 0.71 \\
 19 & Correlation measures and estimation & 72 & 0.71 \\
 14 & Bayesian analysis & 108 & 0.70 \\
 25 & Distribution approximation methods & 100  & 0.67 \\
  5 & Smoothing splines & 51 & 0.66 \\
  8 & Time series & 196 & 0.64 \\
 10 & Regression models & 221 & 0.64 \\
 21 & Density estimation & 65 & 0.64 \\
  1 & Treatment effects & 91 & 0.62 \\
  3 & Confidence intervals & 88 & 0.62 \\
 22 & Censored data and survival analysis & 120 & 0.62 \\
 11 & Variance estimation & 103 & 0.61 \\
  6 & Estimation algorithms: EM, generalizations and MCMC & 56 & 0.60 \\
 16 & Missing data & 69 & 0.58 \\
 17 & Comparison and selection of populations & 94 & 0.58 \\
  7 & Mean square error of estimators & 118 & 0.52 \\
 20 & Conditional inference & 91 & 0.50 \\
  9 & Covariance matrix estimation & 128 & 0.47 \\
 13 & Classification methods & 64 & 0.46 \\
  \hline
\end{tabular}
\end{table}

\begin{table}
\caption{Top 25 leading topics published in the \emph{Journal of the American Statistical Association}, from 1970 to 2015.} \label{tab:jasa}

\centering \footnotesize
\begin{tabular}{rlcc}
  \hline
 \multirow{2}*{\textbf{Cluster}} &\multirow{2}*{\textbf{Topic}} & \textbf{No. of} & \textbf{Cohesion}\\
                       &                    &  \textbf{ papers}   &\textbf{ index }\\
    \hline
 15 & Hypothesis testing & 1060 & 0.84 \\
  4 & Forecast methods in applied contexts & 83 & 0.82 \\
  8 & Cluster analysis and model-based clustering & 128 & 0.80 \\
  1 & Rank data analysis & 93 & 0.74 \\
 17 & Missing data imputation & 91 & 0.71 \\
 13 & Design of experiments & 206 & 0.68 \\
 20 & Spatial data analysis & 104 & 0.68 \\
 23 & Confidence intervals & 208  & 0.68 \\
 22 & Contingency tables & 104 & 0.67 \\
  5 & Regression models & 486 & 0.66 \\
 25 & Variable and model selection & 165  & 0.66 \\
  2 & Nonparametric density estimation & 134 & 0.65 \\
  6 & Censored data and survival analysis & 117 & 0.65 \\
 24 & Measurement errors & 270 & 0.63 \\
 18 & Variance estimation & 185 & 0.62 \\
  3 & Bayesian analysis & 306 & 0.61 \\
 19 & Time series & 408 & 0.61 \\
  7 & Sampling theory & 190 & 0.60 \\
  9 & Smoothing methods & 118 & 0.59 \\
 14 & Distribution approximations & 119 & 0.59 \\
 16 & Treatment and causal effects & 303 & 0.58 \\
 21 & Maximum likelihood & 258 & 0.57 \\
 11 & Network and graphical models & 167 & 0.44 \\
 10 & Robust estimation & 333 & 0.37 \\
 12 & Risk rate estimation & 273 & 0.36 \\
  \hline
\end{tabular}
\end{table}

\begin{table}
\caption{Top 25 leading topics published in the \emph{Journal of the Royal Statistical Society: Series B}, from 1970 to 2015.} \label{tab:jrssb}

\centering \footnotesize
\begin{tabular}{rlcc}
  \hline
 \multirow{2}*{\textbf{Cluster}} &\multirow{2}*{\textbf{Topic}} & \textbf{No. of} & \textbf{Cohesion}\\
                       &                    &  \textbf{ papers}   &\textbf{ index }\\
    \hline
  6 & Bootstrap methods & 34 & 0.84 \\
 14 & Design of experiments & 156 & 0.84 \\
 15 & Hypothesis testing & 229 & 0.84 \\
  2 & Extreme value analysis & 31 & 0.78 \\
 18 & Spatial statistics & 49 & 0.77 \\
 19 & Confidence intervals & 53 & 0.73 \\
 24 & Smoothing methods & 46 & 0.73 \\
 21 & Maximum likelihood & 163 & 0.71 \\
  7 & Priors in Bayesian analysis & 65 & 0.68 \\
 11 & Variable selection & 30 & 0.68 \\
  9 & Regression models & 131 & 0.66 \\
  3 & Bayesian analysis & 88 & 0.65 \\
 23 & Model selection methods and criteria & 37  & 0.65 \\
 17 & Treatment and causal effects & 63 & 0.64 \\
 22 & Time series & 152 & 0.64 \\
 25 & Density estimation: semiparametric and nonparametric & 63  & 0.64 \\
  8 & EM and MCMC algorithm & 84 & 0.63 \\
 16 & Distribution approximations (saddlepoint, ...) & 52 & 0.63 \\
 10 & Measurement error problems & 66 & 0.60 \\
  1 & Generalized linear models & 91 & 0.59 \\
  4 & Dimension reduction and variable selection & 50 & 0.57 \\
  5 & Sampling theory & 44 & 0.57 \\
 12 & False discovery rate & 39 & 0.54 \\
 13 & Association and conditional independence & 52 & 0.46 \\
 20 & Multivariate normality assessment and extensions & 82 & 0.45 \\
  \hline
\end{tabular}
\end{table}

\begin{table}
\caption{Top 25 leading topics published in \emph{Statistical Science}, from 1970 to 2015.} \label{tab:ss}

\centering \footnotesize
\begin{tabular}{rlcc}
  \hline
 \multirow{2}*{\textbf{Cluster}} &\multirow{2}*{\textbf{Topic}} & \textbf{No. of} & \textbf{Cohesion}\\
                       &                    &  \textbf{ papers}   &\textbf{ index }\\
    \hline
  6 & Bootstrap methodology & 18 & 0.96 \\
  2 & Network modelling & 18 & 0.85 \\
 23 & Confidence intervals & 7  & 0.84 \\
  7 & R. A. Fisher & 20 & 0.83 \\
 22 & Causal inference & 19 & 0.82 \\
 14 & Hypothesis testing & 55 & 0.81 \\
 20 & Clinical trials & 18 & 0.81 \\
  5 & Graphical models & 18 & 0.78 \\
  1 & Bayesian analysis & 65 & 0.76 \\
 15 & Priors in Bayesian analysis & 20 & 0.76 \\
  3 & Markov Chain Monte Carlo methods & 21 & 0.75 \\
 13 & Design of experiments & 24 & 0.70 \\
 11 & Estimation algorithms & 34 & 0.69 \\
 19 & Random and treatment effects & 27 & 0.68 \\
 25 & Regression models & 41  & 0.67 \\
 24 & Risk assessment problems & 23 & 0.63 \\
  8 & Missing data problems & 16 & 0.61 \\
 10 & Birnbaum Argument and likelihood & 25 & 0.61 \\
 17 & Prediction problems and methods & 26 & 0.61 \\
  4 & Analysis of genetic data and genomics & 14 & 0.58 \\
 21 & Analysis of temporal data: time series and survival analysis & 22  & 0.53 \\
  9 & Spatial statistics & 26 & 0.50 \\
 12 & Genome-wide association studies & 22 & 0.46 \\
 16 & Computing environments for data analysis& 34 & 0.45 \\
 18 & Reviews and statistics historical papers & 38 & 0.34 \\
  \hline
\end{tabular}
\end{table}

\section{Dynamic clustering} \label{sec:dynamic_clustering}
The main statistical research topics naturally born, evolve or die during time. New interesting topics could emerge at some time, they may capture research interest and related developments, and at some point they could diverge into something else or they could disappear. Moreover, the life process of a topic may have a variable length. 

 \cite{proc6} proposed an approach to perform dynamic topic modeling by fitting a Gaussian time series on the natural parameters of the multinomial distributions that represent the topics. The approach aims to analyze the topic evolution and it seems promising in terms of topic prediction; however it does not describe how new topics appear or disappear over time, mainly because it relies on the restrictive assumption of a fixed number of topics over time.

In order to both address this issue and describe the formation and the evolution of the topics, a dynamic clustering strategy based on semi-supervised mixtures is here developed \citep{Ambroise2000,come,zhu,Vandewalle}. We assume a forward evolution perspective, that means documents at the present temporal frame are projected on a future classification frame. Imagine several documents are classified around $k_1$ main topics in a certain temporal interval and a second set of documents is classified around $k_2$ topic clusters in a subsequent temporal interval. We are interested in studying the evolution of the first set of documents towards the future $k_2$ groups. In order to capture the dynamic forward process the first-interval documents can be projected into the second-interval classification and the dynamic is considered `relevant' if the relative fraction of movement is above a specified cutoff.

More formally, suppose different sets of documents $\textbf{y}^{(1)},\ldots,\textbf{y}^{(T)}$, with dimensions $(n_1 \times p_1),\ldots,(n_T \times p_T)$ respectively, are collected in different time spans, say $1,\ldots,t,\ldots,T$. Assume $\textbf{y}^{(t)}$ at time $t$ is clustered into $k_t$ groups and $\textbf{y}^{(t+1)}$ into $k_{t+1}$ groups. Since $p_t\neq p_{t+1}$ we first transform $\textbf{y}^{(t)}$ into a `matched' version $\tilde{\textbf{y}}^{(t)}$ such that it has dimension $n_{t}\times p_{t+1}$ by discarding the `missing in future' columns and artificially adding the eventual new term columns with frequency zero.
Then a semi-supervised version of the mixture model (\ref{eqn:mixture}) is fitted on $\textbf{y}=(\tilde{\textbf{y}}^{(t)},\textbf{y}^{(t+1)})$ with $k_{t+1}$ components and known labels, $\textbf{z}^{(t+1)}$, for $\textbf{y}^{(t+1)}$. The log-likelihood to be maximized with respect to $\boldsymbol\pi$ only is:
\begin{eqnarray}\label{eqn:slogL}
\ell(\boldsymbol\pi;\boldsymbol\xi,\tilde{\textbf{y}}^{(t)},\textbf{y}^{(t+1)},\textbf{z}^{(t+1)})
&=&\sum_{j=1}^{n_t}\log\sum_{i=1}^{k_{t+1}}\pi_i\left( e^{-\lambda d(\tilde{y}_j^{(t)},\xi_i)}\right) \nonumber  \\ &+&\sum_{j=1}^{n_{t+1}}\sum_{i=1}^{k_{t+1}}z_{ji}^{(t+1)}\left(\log \pi_i -\lambda d({y}_j^{(t+1)},\xi_i)\right) \nonumber  \\
&+& (n_t+n_{t+1})\log \psi(\lambda)
\end{eqnarray}

The dynamic parameter estimation can be obtained by an adapted version of the EM-algorithm previously described, where the allocation variable is $\textbf{z}^{(t)}$ is latent and estimated in the E-step, while $\textbf{z}^{(t+1)}$ is known. The precision parameter is taken fixed for all $t$. The M-step for $\boldsymbol\pi$ does not change, while the previously estimated centroid, $\boldsymbol\xi$, are taken as known. This naturally produces a classification of the first set of units $\tilde{\textbf{y}}^{(t)}$ into $k_{t+1}$ groups. The projected new classification, say $\tilde{c}_{t+1}$ can be compared to the classification originally obtained at $t$, ${c}_{t}$. The confusion matrix in Table \ref{tab:tabcont} contains the frequencies of the documents collected at $t$ that are allocated to a certain cluster at time $t$ and $t+1$.

\begin{table}
\small
\caption{Number of statistical papers $\textbf{y}^{(t)}$ classified in $k_t$ and $k_{t+1}$ groups. \label{tab:tabcont}}
\begin{center}
\begin{tabular}{|l|rrrr|}
\hline
\ \ \ \ \ $\tilde{c}_{t+1}$ & 1 & 2 & ... & $k_{t+1}$ \\
$c_t$  & & & & \\ \hline
1 & $n_{11}$ & $n_{12}$& ... & $n_{1k_{t+1}}$\\
2 & $n_{21}$ & $n_{22}$& ... & $n_{2k_{t+1}}$\\
... & ... & ... & ... & ...\\
$k_t$ & $n_{k_t1}$ & $n_{k_t2}$ &... & $n_{k_tk_{t+1}}$\\
\hline
\end{tabular}
\end{center}
\end{table}

By denoting with $n_{uv}, \ u=1,\ldots,k_t; \ v=1,\ldots,k_{t+1}$ the number of papers that are allocated into the group $u$ at time $t$ and in group $v$ at time $t+1$, the relative frequencies ${f_{uv}=\frac{n_{uv}}{\sum_{v=1}^{k_{t+1}} n_{uv}}}$ measure the migration of documents from cluster $u$ to cluster $v$. When $f_{uv}>s$, where $s$ is a cutoff value between 0 and 1, we can reasonably consider $v$ as a pursuance of $u$ and establish a dynamic connection between the two clusters.

\subsection{Analysis and Results}
To study the evolution of the leading statistical topics, we need to firstly cluster the abstracts separately for temporal intervals; the datasets are quite similar in terms of both the number of documents (except for the last period that has fewer abstracts since it extends for six years only) and the number of words. In order to reduce the dimensionality, a variable selection is performed, similarly to that described in section \ref{sec:static_clustering}. Words are sorted according to their entropy and in order to assure a comparable dimension problem over time, the first 700 words are retained: this allows to consider terms with, on average, an entropy of at least 0.40. Table \ref{tab:words_time} reports the number of terms for each interval with (first row) and without (second row) a 0.40 entropy threshold.

\begin{table}
\small
\caption{Number of words for each time span, with and without a threshold on entropy value $H$.} \label{tab:words_time}
\begin{center}
\begin{tabular}{lrrrrrr}
\hline \textbf{No. of words}   & \textbf{1970-1979}  & \textbf{1980-1989}   & \textbf{1990-1999}    & \textbf{2000-2009}    & \textbf{2010-2015}\\
\hline Unconstrained $H$       & 5148         &6525   &   7329        &7907    &	6339\\
$H\geq$0.40     & 536          & 642   &	798         &860     &	774\\
\hline
\end{tabular}
\end{center}
\end{table}

The number of topics that have led each decade can be different, according to the themes that characterized the intervals; in order to fully and properly describe the temporal evolution, several models with different number of clusters are estimated on each dataset. In particular, we run the EM algorithm of section \ref{subsec:modelestimation}, allowing for a number of clusters $k$ varying from 2 to 20.

In principle, the number of clusters can be selected according to classical information criteria such as the Bayesian (BIC) and the Akaike (AIC) information criteria. However, since the normalization constant is unknown the conventional criteria cannot be used to compare models characterized by different $\lambda$s.
In order to overcome this limit and, therefore, to take advantage of the information criteria for model selection, we estimate a single precision parameter by averaging multiple $\lambda$s. In particular, we run several $k$-means algorithm with an increasing number of groups, say from 2 to 20; for each classification $\hat{\lambda}$ is estimated according to equation (\ref{eqn:rel}), by considering $\alpha=0.05$. The precision parameter to be used in the dynamic clustering for each dataset is, therefore, obtained by computing the mean of the 19 values. The classification and the cluster prototypes from the $k$-means clustering are used as initial values for the EM algorithm.


Once the documents of each interval are classified according to the most appropriate number of groups (according to the AIC they are 16,16,18,20 and 15 in the five time intervals respectively), the dynamic clustering described in section \ref{sec:dynamic_clustering} can be performed. Group characterization is then possible by studying the most frequent words and the abstracts closer to prototypes.

Figure \ref{fig: dynamic} represents the dynamic clustering. For each time span, displayed in columns, the corresponding clusters are plotted; if a topic persists in the following time span, an arrow will link that group with its subsequent. Topics may originate and disappear in the same decade, or the may evolve into something different. For a topic to survive it needs to have at least the 40\% of its abstracts projected into a single cluster. Dashed arrows link groups whose relative fraction of movement is between 0.40 and 0.70, whereas solid arrows link groups whose percentage of projected documents is larger than 70\%.


\afterpage{%

\thispagestyle{empty}
\begin{figure}[p]
  \centering
\includegraphics[width=\textwidth]{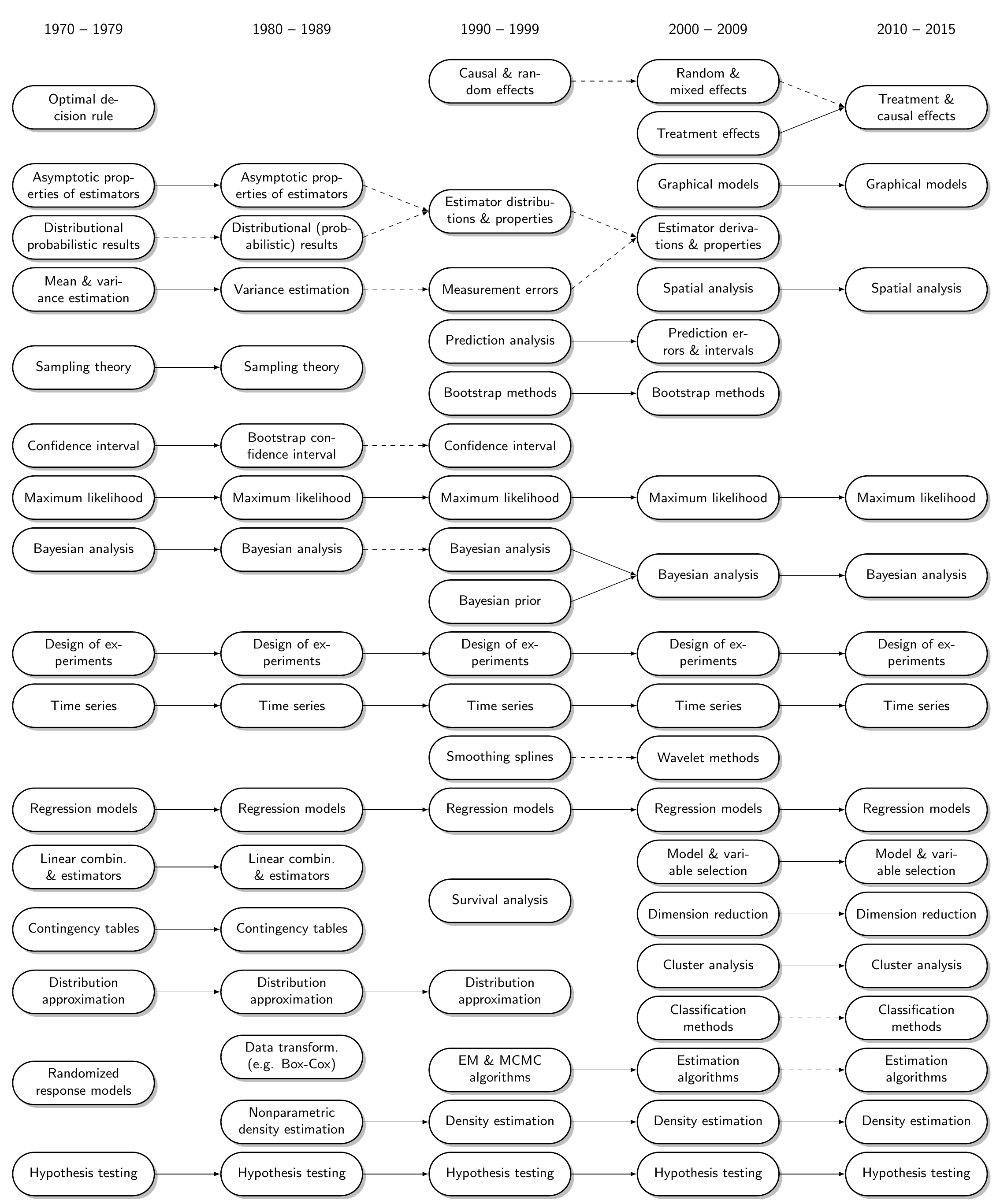}
  \caption{Dynamic clustering of abstracts from 1970 to 2015. Dashed arrows link groups whose relative fraction of movement is between 0.40 and 0.70, solid arrows between 0.70 and 1.}\label{fig: dynamic}
\end{figure}
\clearpage
}

Some topics are evergreen: \emph{hypothesis testing}, \emph{time series}, \emph{design of experiments}, \emph{maximum likelihood}, \emph{regression models} are themes that, despite having changed quite a lot in the past, have never lost their centrality in the statistical literature since 1970. \emph{Bayesian analysis} covered the whole period as well; in the decade 1980-1989, a growing interest allowed to obtain a separate cluster on the \emph{Bayesian prior}, that afterward converged to the general Bayesian one.

Differently, \emph{confidence intervals} was a leading topic for three decades, with a special attention to \emph{bootstrap confidence interval} between 1980 and 1989. After that, \emph{bootstrap methods} loomed out as a separate cluster.

Articles on \emph{random effects} became particularly numerous since the decade 1990-1999, incorporating later also mixed and treatment effects. The estimation problem has been very relevant as well: at the very beginning there were three groups about estimation, including a group dedicated only to the \emph{mean and variance estimation}. A separate flow is reserved to the \emph{nonparametric literature}, that emerged in both density estimation and splines contexts.

The dynamic clustering proved to be able to identify the moment when topics like \emph{bootstrap} and \emph{estimation algorithms} (like EM or MCMC) became more popular: although the first theoretical contributions on these topics arose between 70s-80s, the scientific production sprang later, from the 90s, helped by an increasing availability of computing machines.

Similarly, the development in many other areas of scientific research of the last twenty years led to a huge availability of \emph{big data}; this translates to the urge of statistical tools able to deal with such data. In fact, classical multivariate methods found a renewed impulse: from the decade 2000-2009 topics like \emph{cluster analysis}, \emph{spatial analysis}, \emph{dimension reduction}, \emph{model and variable selection} and  \emph{classification methods} are expressed as separate trends, highlighting their importance in the recent statistical literature.

\section{Conclusions}
We presented a mixture of cosine distances that allowed us to identify and to describe the 25 most important topics published in five prestigious journals from 1970. The data were collected by considering the title and the abstract of each statistical article published in the Annals of Statistics, Biometrika, Journal of American Statistical Association, Journal of the Royal Statistical Society, series B and Statistical Science. 


The detected twenty-five clusters have different size and cohesion, according to the degree of heterogeneity and generality of each topic. In so doing we obtained a sort of taxonomy of the main statistical research themes discussed in the last forty-five years. In addition, we zoomed in on the considered journals: clustering the papers separately for the different journals allowed to distinguish the transversal topics from more specific themes, that resulted in a smaller set of periodicals.

Each classification is a simplification of the reality. Each summary brings new knowledge but, at the same time, implies losses. Our taxonomy is not an exhaustive list of the interconnected topics of the recent statistical research: some of the obtained groups are in fact very general (e.g. maximum likelihood, graphical models, regression models, ...) and include a variety of sub-themes that are relevant and interesting, but not important enough to be uncluttered to form separate clusters (e.g. E-M and MCMC algorithms joined the larger group of \emph{estimation algorithms}).

Since the main statistical research topics naturally born, evolve or die during time, we also developed a dynamic clustering strategy that allowed to follow the projection of a statistical theme in the following decades. Data were organized in time intervals (1970-79, 1980-89, 1990-99, 2000-09, 2010-15). Each period was characterized by a different number of groups, chosen via AIC from a sequence ranging from 2 to 20. Our approach did not aim to spot the true introduction of a statistical topic, rather to detect the moment when a certain theme became `popular' and `trendy' , how it evolved, and also its `decadence'. Our dynamic clustering strategy is based on semi-supervised mixtures and mutual comparisons between the actual static classification and the predicted dynamic one in a forward perspective. The idea is pretty simple but really effective in detecting the topic evolution together with the description of how new topics appear or disappear over time.

\bibliographystyle{Chicago}

\bibliography{ref}

\end{document}